\documentclass[a4paper]{jpconf}
\usepackage{graphicx}
\usepackage{amsmath,amssymb}
\usepackage{hyperref}
\usepackage{epstopdf}
\usepackage{bm}
\usepackage{color}

\def\em{{\rm em}}

\renewcommand{\emph}[1]{\textit{#1}}

\begin{document}
\title{Axion Dark Matter Search with Interferometric Gravitational Wave Detectors}

\author{Koji Nagano$^1$, Ippei Obata$^1$, Tomohiro Fujita$^2$, Yuta Michimura$^3$}
\address{$^1$Institute for Cosmic Ray Research, University of Tokyo, Kashiwa 277-8582, Japan, \\
$^2$Department of Physics, Kyoto University, Kyoto, 606-8502, Japan, \\
$^3$Department of Physics, University of Tokyo, Bunkyo, Tokyo 113-0033, Japan}
\ead{obata@tap.u-tokyo.ac.jp}

\begin{abstract}
Axion dark matter differentiates the phase velocities of the circular-polarized photons.
In [{\it Phys. Rev. Lett.\/} \href{https://doi.org/10.1103/PhysRevLett.123.111301}{{\bf 123}, 111301 (2019)}], we have proposed a scheme to measure the phase difference by using a linear optical 
cavity. If the scheme is applied to the Fabry-P\'erot arm of Advanced LIGO-like (Cosmic-Explorer-like) gravitational wave
detector, the potential sensitivity to the axion-photon coupling constant, $g_{\text{a}\gamma}$, reaches $g_{\text{a}\gamma} \simeq 8\times10^{-13} 
\text{~GeV}^{-1}\, (4 \times 10^{-14}\text{~GeV}^{-1})$  
at the axion mass $m \simeq 3\times 10^{-13}$ eV ($2\times10^{-15}$ eV)
and remains at around this sensitivity for 3 orders of magnitude in mass.
Furthermore, its sensitivity has a sharp peak reaching 
$g_{\text{a}\gamma} \simeq 10^{-14} \text{~GeV}^{-1}\  (8\times10^{-17} \text{~GeV}^{-1})$ 
at $m = 1.563\times10^{-10}$ eV ($1.563\times10^{-11}$ eV).
This sensitivity can be achieved without loosing any sensitivity to gravitational waves.\end{abstract}

\section{Introduction}
Axion, a pseudo-scalar field motivated by QCD or string theory, typically has a small mass $m \ll \text{eV}$ 
and behaves like non-relativistic fluid in the present universe due to its oscillatory behavior: $a(t) = a_0\cos(mt + \delta_\tau(t))$ with its constant amplitude $a_0$ and a phase factor $\delta_\tau(t)$.
For this reason, axion is a cosmologically well-motivated candidate of dark matter.
Another important feature is that axion is topologically coupled to photon $g_{\text{a}\gamma}a(t) F_{\mu \nu} \tilde{F}^{\mu \nu}$ via the coupling constant $g_{\text{a}\gamma}$.
The conventional way to probe axion is to look for a phenomenon that axion and photon are converted each other under the background magnetic field, known as the axion-photon conversion.
Many experiments and astronomical observations have been performed to probe axion  via the axion-photon conversion, while no signal has been found \cite{Irastorza2018}.

Recently, however, a new experimental approach to search for axion dark matter was proposed which does not need a strong magnetic field but uses optical cavity~\cite{Melissinos2009, DeRocco2018, Obata2018, Liu:2018icu}.
This new method aims to measure the difference of phase velocity between two circular-polarized photons $\delta c(t) 
= |c_L-c_R| = 2\delta c_0 \sin(m t + \delta_\tau(t))$ which is caused by the coupling to axion dark matter
\begin{align}
\delta c_0 \simeq 1.3 \times 10^{-24} \left( \frac{\lambda}{1550 \text{ nm}}\right) 
\left( \frac{g_{\text{a}\gamma}}{10^{-12} \text{ GeV}^{-1}}\right)
\end{align}
with the frequency of axion mass $f = m/(2\pi) \simeq 2.4 ~\text{Hz}~(m/10^{-14}~\text{eV})$.
Here we assumed the laser light with a wavelength $\lambda = 2\pi/k$ and used the present energy density value of axion dark matter around earth, $\rho_a = a_0^2m^2/2 \simeq 0.3 \text{ GeV/cm}^3$.
The experimental sensitivity is only limited by quantum noise in principle and it can probe tiny axion-photon coupling $g_{\text{a}\gamma} \lesssim 10^{-11} ~\text{GeV}^{-1}$ with axion mass range $m \lesssim 10^{-10} ~\text{eV}$ which is competitive with other experimental proposals.
Moreover, this new method can be highly advantageous compared with the conventional axion detectors, since it does not require superconducting magnets which often drive large cost.
Therefore, we expect that this method opens a new window to the axion dark matter research.

Inspired by these proposals using optical cavity, we propose a new scheme to search for axion dark matter by using a large linear Fabry-P\'erot cavity such as gravitational wave detectors \cite{Nagano:2019rbw}.
Remarkably, our new method enables the interferometers to probe axion-like dark matter during the gravitational wave observation run without loosing any sensitivity to gravitational waves.
We estimate the potential sensitivity to $g_{\text{a}\gamma}$ with the parameter sets of gravitational wave observatories. 
Their sensitivities can overcome the current upper limit with broad axion mass range and put better bounds than the proposed axion experiments.

\section{Axion search with a linear optical cavity}

\begin{figure}[htb]
\begin{center}
\begin{tabular}{c}
\begin{minipage}{0.5\hsize}
\begin{center}
\includegraphics[width=0.95\textwidth]{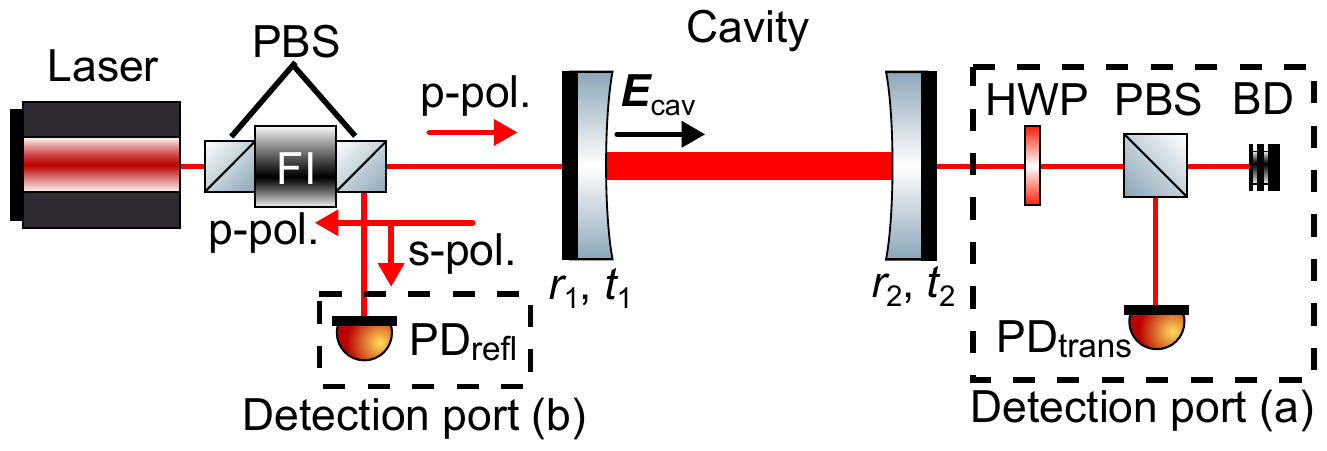}
\end{center}
\end{minipage}
\begin{minipage}{0.5\hsize}
\begin{center}
\caption{Schematic of experimental setup for axion search with a linear optical cavity. 
  FI, Faraday isolator; HWP, half wave plate; PBS, polarizing beam splitter; 
  PD, photodetector; BD, beam dump.
Components for phase measurement are not shown.
}
\label{fig:Setup22}
\end{center}
 \end{minipage}
\end{tabular}
\end{center}
\end{figure}

We present how to detect the modulation of speed of light with linear optical cavities.
The key point is that one linearly polarized light (e.g. horizontal polarization, that is p-polarization) is polarization-modulated due to axion dark matter and the orthogonally polarized light
(e.g. vertical polarization, that is s-polarization) is produced because the linearly polarized light can be expressed by a superposition of two circularly polarized lights.
The schematic setup of our proposed scheme is shown in figure \ref{fig:Setup22}.
First, as a carrier wave, we input linearly-polarized monochromatic laser light with the angular frequency which corresponds to the wave number $k$.
Here, we consider p-polarized light as input light without loosing generality.
The cavity consists of the input and output mirrors whose amplitude reflectivities and transmissivities are represented by ($r_1$, $t_1$) and ($r_2$, $t_2$).
When the cavity is kept to resonate with a phase measurement,
the beam is accumulated inside cavity.
After the calculation, the electronic field in the cavity is written as, 
\begin{align}
\bm{E}_\text{cav}(t) 
=& \frac{t_1}{1-r_1r_2}\left[\bm{E}^\text{p}(t) 
- \delta \phi(t)\bm{E}^\text{s}(t)\right],
\end{align}
where $\bm{E}^\text{p/s}$ are electric vectors of p/s-polarized light,
$\delta \phi(t) \equiv \int^\infty_{-\infty} 
 \tilde{\delta c}(m) H_\text{a}(m) e^{im t} dm/2\pi$ is a signal
caused by axion dark matter with a response function of cavity 
\begin{equation}
 H_\text{a}(m) \equiv i \frac{k}{m} 
 \frac{4 r_1r_2\sin^2\left(\frac{m L}{2}\right)}
 {1 - r_1 r_2 e^{-i2m L}}  \left(-e^{-imL}\right) \label{eq:Ha}.
\end{equation}
Here, we treat $\delta_\tau(t)$  as constant since we only consider the axion mass range where the axion oscillation coherent time ($\tau \sim 1(10^{-16}\text{eV}/m)[\text{year}]$) is sufficiently longer than the storage time of the optical cavity.
Equation (\ref{eq:Ha}) indicates that the signal is enhanced in proportion to 
$r_1 r_2/(1-r_1 r_2)$ at 
$m = \pi/L$, which corresponds to the free spectral range.
In addition, $H_\text{a}(m) \propto 1/m$ at $mL= \pi(2N-1) \ (N\in\mathbb{N})$ since the axion effect on the photons in the cavity is cancelled out except for the last half of the axion oscillation when the axion oscillation period is shorter than the photon storage time of the cavity. 
In low mass range ($mL \ll 1$), $H_\text{a}(m) \propto m$ since the axion effect is cancelled on going and returning way.
Then the signal is detected in detection port (a) or (b) as polarization modulation with polarizing optics. 
In detection port (a), the polarization of transmitted light from the cavity is slightly 
rotated by the half wave plate. Then, the photodetector ($\text{PD}_\text{trans}$) 
receives  s-polarized light generated by axion-photon coupling as a beatnote with faint (but much stronger than the signal) carrier wave, while most of the carrier light is transmitted by the polarizing beam splitter (PBS).
In detection port (b), the PD ($\text{PD}_\text{refl}$) receives signal reflected by the faraday isolator (FI) as a beatnote with faint carrier wave again.
%
%
In this case, the carrier wave is generated by non-ideal birefringence between the cavity and FI, such as input mirror substrate.
These two detection ports can be added without modifying the instrument for the phase measurement.

\section{Sensitivity to the axion-photon coupling}

We estimate the potential sensitivity of the linear cavity to axion-photon coupling.
Here, only shot noise which is caused by vacuum fluctuation of electric field is considered in a similar way to shot-noise estimation of gravitational wave detectors~\cite{Kimble2001a}.
The one-sided linear spectrum of shot noise equivalent to $\tilde{\delta c}(m)$, 
$\sqrt{S_\text{shot}(m)}$, is obtained by considering the ratio of the noise term to the signal term,
\begin{equation}
\sqrt{S_\text{shot}(m)} = \frac{\sqrt{k/(2 P_0)}}
{\sqrt{\mathcal{T}_j}|H_\text{a}(m)|}, \qquad \sqrt{\mathcal{T}_j} \equiv \frac{t_1t_j}{1-r_1r_2} \quad (j=1,2)
\label{eq:Shotnoise}
\end{equation}
where $P_0$ is incident power and $j = 2 \ (1)$ represents the detection port (a) ((b)).
According to the equation (\ref{eq:Shotnoise}), if the cavity is over-coupled, i.e. $t_1 > t_2$, 
detection port (b) is better.
On the other hand, detection port (a) is effective for the critical-coupled cavity, i.e. $t_1 = t_2$, since there is no carrier wave in the reflection port under the critical coupling condition.
If the sensitivity is limited by shot noise, the signal-to-noise ratio (SNR) for $\delta c_0$ is improved with measurement time, $T_\text{obs}$. The improvement 
depends on whether $T_\text{obs}$ is larger than the coherent time of axion oscillation, $\tau$, or not \cite{Budker2014}:
\begin{equation}
\text{SNR} = 
  \begin{cases}
    \frac{\sqrt{T_\text{obs}}}{2\sqrt{S_\text{shot}(m)}}\delta c_0
    & (T_\text{obs} \lesssim \tau) \\
    \frac{(T_\text{obs} \tau)^{1/4}}{2\sqrt{S_\text{shot}(m)}}\delta c_0
    & (T_\text{obs} \gtrsim \tau)
  \end{cases} .
\end{equation}
We can find the detectable value of $\tilde{\delta c}(m)$ which sets the SNR to unity.
Finally, this is translated into the sensitivity to $g_{\text{a}\gamma}$.
\begin{figure}[htb]
  \centering
  \includegraphics[width=4in]{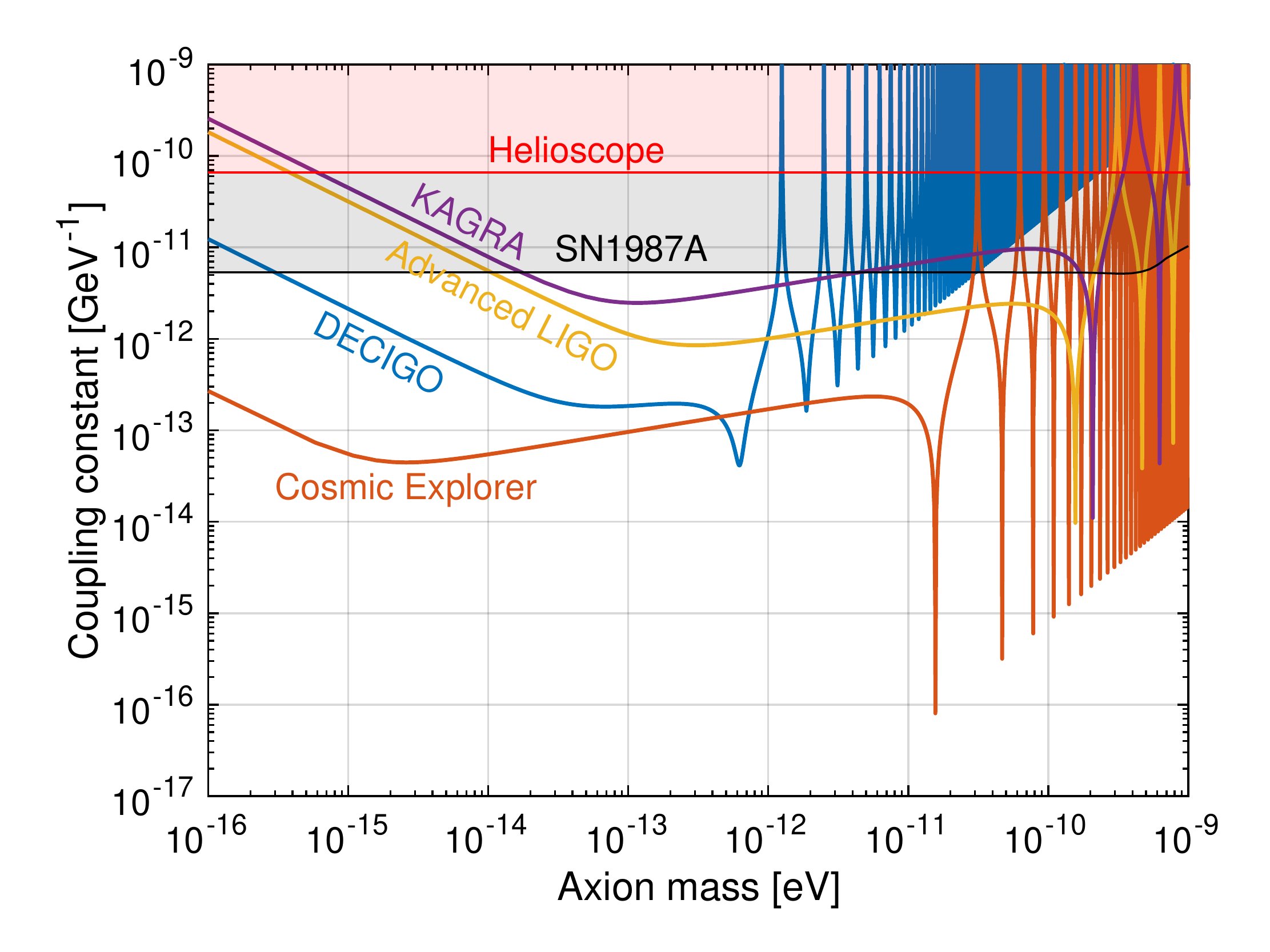}
  \caption{The sensitivity curves for $g_{\text{a}\gamma}$ with respect to the axion mass in comparison with several interferometric detectors.
  The red and gray band express the current limit provided by CAST~\cite{CASTCollaboration2017} and the cosmic ray observations of SN1987A~\cite{Payez2015}.
  }
  \label{fig:SensComp}
\end{figure}

Figure \ref{fig:SensComp} shows the shot-noise limited sensitivities 
to $g_{\text{a}\gamma}$ with our scheme.
Here, we adopted the experimental parameter sets used or planned by gravitational wave detectors (specifically, DECIGO~\cite{Kawamura2008}, 
CE~\cite{Abbott2017}, KAGRA~\cite{Somiya2012}, and aLIGO~\cite{Aasi2015}).
We also assume $T_\text{obs} = 1$ year and $r_i^2 + t_i^2 = 1$.
Note that detection port (a) is used for DECIGO-like 
detector and port (b) is used for CE-, KAGRA-, and aLIGO-like detectors.
In all parameter sets, the upper limit provided by CAST ~\cite{CASTCollaboration2017} can be improved.
Especially, the CE-like detector can overcome the CAST limit by three 
orders of magnitude in broad mass range around between 
$4\times10^{-16}$ and $1\times10^{-13}$ eV.
%
%
It is worth noting that in our scheme the displacement noise such as the vibration of mirrors or the gravitational wave signal itself does not become manifest unlike a gravitational wave detector.
This is because the displacement noises and gravitational waves make the same phase shift in the two circularly-polarized lights propagating in the same path 
and this phase shift is cancelled in the measurement of the phase difference between two polarized lights.
A major technical noise source in our scheme is a roll motion of the mirrors which would generate relative phase shift in the two polarized lights through birefringence of the mirror coating. 
We have checked that the noise spectrum in aLIGO case is determined by $\sqrt{S_\text{roll}} < 3\times10^{-26}$ 1/Hz$^{1/2}$ for $m>10^{-14}$ eV, which is smaller than shot noise level \cite{Nagano:2019rbw}. 
In DECIGO, CE and KAGRA, the roll motion of the mirror would be small since they would be in space or underground site while aLIGO is on the ground.
For more realistic discussion, however, we will come to need the measured values of birefringence actually used in each interferometer experiment.

In order to apply our method to the real gravitational detector, some optics are
added for detection port and there exist constructional problems.
The approach to detect the signal in detection port (b) is not quite simple because there have been equipped several apparatuses, such as a beam splitter, a signal recycling mirror, and so on, between the front mirror and FI.
In principle, the axion signal can be extracted behind the signal recycling mirror as with 
the gravitational wave signal readout.
More practical issues will be investigated in future work. 



\ack
In this work, KN, IO, YM and TF are supported by the JSPS KAKENHI Grant No.~JP17J01176, JSPS KAKENHI Grant No.~19K14702, JSPS Grant-in-Aid for Scientific Research (B) No.~18H01224 and Grant-in-Aid for JSPS Research Fellow No.~17J09103, respectively.


\section*{References}
\begin{thebibliography}{99}


\bibitem{Irastorza2018}
I.~G. Irastorza and J.~Redondo,
\newblock Progress in Particle and Nuclear Physics {\bf 102}, 89 (2018);
%

\bibitem{Melissinos2009}
A.~C. Melissinos,
\newblock Phys. Rev. Lett. {\bf 102}, 202001 (2009).

\bibitem{DeRocco2018}
W.~DeRocco and A.~Hook,
\newblock Phys. Rev. D {\bf 98}, 035021 (2018).

\bibitem{Obata2018}
I.~Obata, T.~Fujita, and Y.~Michimura,
\newblock Phys. Rev. Lett. {\bf 121}, 161301 (2018).

\bibitem{Liu:2018icu} 
  H.~Liu, B.~D.~Elwood, M.~Evans and J.~Thaler,
\newblock Phys.\ Rev.\ D {\bf 100}, no. 2, 023548 (2019).


\bibitem{Nagano:2019rbw} 
  K.~Nagano, T.~Fujita, Y.~Michimura and I.~Obata,
\newblock Phys.\ Rev.\ Lett.\  {\bf 123}, no. 11, 111301 (2019).

\bibitem{Kimble2001a}
H.~J. Kimble, Y.~Levin, A.~B. Matsko, K.~S. Thorne, and S.~P. Vyatchanin,
\newblock Phys. Rev. D {\bf 65}, 022002 (2001).

\bibitem{Budker2014}
D.~Budker, P.~W. Graham, M.~Ledbetter, S.~Rajendran, and A.~O. Sushkov,
\newblock Phys. Rev. X {\bf 4}, 021030 (2014).

\bibitem{Kawamura2008}
S.~Kawamura et~al.,
\newblock J. Phys.: Conf. Ser. {\bf 120}, 032004 (2008).

\bibitem{Abbott2017}
B.~P. Abbott et~al.,
\newblock Class. Quantum Grav. {\bf 34}, 044001 (2017).

\bibitem{Somiya2012}
K.~Somiya,
\newblock Class. Quantum Grav. {\bf 29}, 124007 (2012);
%
Y.~Aso et~al.,
\newblock Phys. Rev. D {\bf 88}, 043007 (2013);
%
T.~Akutsu et~al.,
\newblock Prog Theor Exp Phys {\bf 2018}, 013F01 (2018).

\bibitem{Aasi2015}
J. Aasi et~al.,
\newblock Class. Quantum Grav. {\bf 32}, 074001 (2015).

\bibitem{CASTCollaboration2017}
V.~Anastassopoulos et~al. [CAST Collaboration],
\newblock Nature Physics {\bf 13}, 584 (2017).


\bibitem{Payez2015}
A.~Payez et~al.,
\newblock J. Cosmol. Astropart. Phys. 02 (2015) 006.



\end{thebibliography}
\end{document}